\documentclass[12pt]{article}

\catcode`\@=11

\global\arraycolsep=2pt
\usepackage{amsbsy,amssymb,latexsym,amsfonts,amsmath}
\usepackage{graphicx,color}

\begin{document}

\begin{flushright}
\parbox{4.2cm}

YITP-23-119
\end{flushright}

\vspace*{0.7cm}

\begin{center}
{ \Large  Functional renormalization group approach to dipolar fixed point which is scale-invariant  but non-conformal}
\vspace*{1.5cm}\\
{Yu Nakayama}
\end{center}
\vspace*{1.0cm}
\begin{center}

Department of Physics, Rikkyo University, Toshima, Tokyo 171-8501, Japan

Yukawa Institute for Theoretical Physics,
Kyoto University, Kitashirakawa Oiwakecho, Sakyo-ku, Kyoto 606-8502, Japan

\vspace{3.8cm}
\end{center}

\begin{abstract}
A dipolar fixed point introduced by Aharony and Fisher is a physical example of interacting scale-invariant but non-conformal field theories. We find that the perturbative critical exponents computed in $\epsilon$ expansions violate the conformal bootstrap bound. We formulate the functional renormalization group equations a la Wetterich and Polchinski to study the fixed point. We present some results in three dimensions within (uncontrolled) local potential approximations (with or without perturbative anomalous dimensions). 
\end{abstract}

\thispagestyle{empty} 

\setcounter{page}{0}

\newpage

\section{Introduction}
Conformal invariance has played a central role in understanding critical phenomena not only in two dimensions but also in higher dimensions. For instance, the conformal invariance is powerful enough to determine the critical exponents of the three-dimensional Ising model in six digits by using the recently developed numerical conformal bootstrap method \cite{ElShowk:2012ht}\cite{El-Showk:2014dwa}\cite{Kos:2014bka}\cite{Simmons-Duffin:2016wlq}\cite{Kos:2016ysd}. There are many other critical phenomena studied by using the conformal bootstrap (see e.g. \cite{Poland:2018epd} for a review).

While powerful enough, it seems mysterious that the critical phenomena show enhanced conformal symmetry rather than mere scale invariance. It is indeed quite challenging to prove that the Ising model at criticality shows conformal invariance. On the other hand, it seems surprisingly hard to find examples of scale-invariant but not conformal field theories in theory \cite{Riva:2005gd}\cite{ElShowk:2011gz}\cite{Nakayama:2016cyh}, let alone in physical examples (see e.g. \cite{Nakayama:2013is} for a review). 

In \cite{R}, it was discussed that (isotropic) dipolar magnet \cite{AF} is one of such rare examples of interacting scale but not conformal field theories.\footnote{Subsequently, the details of this theory including the non-renormalization property of the virial current from the (hidden) shift symmetry is developed in \cite{Gimenez-Grau:2023lpz}.} Because it is not conformal invariant, we cannot use the numerical conformal bootstrap method to investigate its critical exponents. Indeed, in this paper, we will show that the perturbative critical exponents computed in $\epsilon$ expansions violate the conformal bootstrap bound.

With this situation in mind, we investigate the functional renormalization group approaches to the dipolar fixed point. The functional renormalization group is regarded as a non-perturbative method to study the renormalization group flow and its fixed point (see e.g. \cite{Gies:2006wv}\cite{Delamotte:2007pf}\cite{Dupuis:2020fhh} for reviews). Since it does not rely on the conformal symmetry unlike the conformal bootstrap method, it can be applied to the dipolar magnet. In this paper, we use the Wetterich equation \cite{Wetterich:1992yh} as well as the Polchinski equation \cite{Polchinski:1983gv} to investigate the dipolar fixed point. 
We first show that both approaches reproduce the lowest order $\epsilon$ expansions in the local potential approximation with the perturbative truncation. We then present some (non-perturbative) results in three dimensions within (uncontrolled) local potential approximations. 

\section{Functional renormalization group approaches to dipolar fixed point}

\subsection{Dipolar fixed point and violation of bootstrap bound}
In the Landau-Ginzburg description, the Heisenberg magnet in $d$-dimension is described by the effective action
\begin{align}
 S = \int d^d x \left( \frac{1}{2}\partial_\mu \phi_i \partial_\mu \phi_i + t \phi_i^2 + \lambda (\phi_i^2)^2 \right) \ , 
\end{align}
where $i=1,\cdots, d$. It has the global $O(d)$ symmetry (as well as the $O(d)$ spatial rotational symmetry) since the exchange interaction relevant to the Heisenberg magnet only acts on internal spin rather than the orbital spin.\footnote{Strictly speaking, the magnetization is not a ``vector" in $d\neq 3$ dimensions (rather it is a two-form), but we will continue the dimensionality here in order to set up a simple $\epsilon$ expansion.} The renormalization group fixed point of this effective action describes the critical behavior of the Heisenberg magnet. 

A dipolar interaction breaks the separation of the spin rotation and the orbital rotation, resulting in the explicit symmetry breaking of $O(d) \times O(d)$ down to $O(d)$. In the Landau-Ginzburg description, it is described by the effective action
\begin{align}
S = \int d^d x \left( \frac{1}{2}\partial_\mu \phi_\nu \partial_\mu \phi_\nu + \xi (\partial_\mu \phi_\mu)^2 + t \phi_\mu^2 + \lambda (\phi_\mu^2)^2 \right) \ ,\end{align}
We will assume $\xi = \infty$ so that the vector $\phi_\mu$ is purely transverse.\footnote{Within perturbative $\epsilon$ expansions, it turns out that $\xi = \infty$ is an unstable IR fixed point, but there is a (hidden) symmetry that makes it possible to set $\xi=\infty$ under the renormalization group flow. See \cite{Gimenez-Grau:2023lpz} for a complete analysis of the story.}
Alternatively one may use the Lagrange multiplier formulation
\begin{align}
S = \int d^d x \left( \frac{1}{2}\partial_\mu \phi_\nu \partial_\mu \phi_\nu + U \partial_\mu \phi_\mu  + t \phi_\mu^2 + \lambda (\phi_\mu^2)^2 \right) \ ,\end{align}
where $U$ is the Lagrange multiplier. 
In this picture, it is easier to see that the transverse condition is not renormalized because of the shift symmetry of $U$. The critical behavior of the dipolar magnet is described by the renormalization group fixed point of this action.

Aharony and Fisher did the perturbative studies of the renormalization group flow in $d=4-\epsilon$ dimensions. We quote their results \cite{AF}\cite{AF2}\cite{AF3} (see also \cite{Kudlis} for three loop results directly in three dimensions). The scaling dimension of the lowest non-trivial singlet operator $\Delta_t$ is given by
\begin{align}
\Delta_t = 2-\frac{8}{17}\epsilon \ .
\end{align}
The scaling dimension of the lowest vector operator $\Delta_{\phi}$ is given by
\begin{align}
\Delta_{\phi} = \frac{2-\epsilon}{2} + \frac{10}{867} \epsilon^2 \ . 
\end{align}
In comparison, let us also quote the scaling dimensions of the corresponding operator in the critical $O(N)$ model
\begin{align}
\Delta_t &= 2 - \frac{6}{N+8}\epsilon \cr
\Delta_\phi & = \frac{2-\epsilon}{2} + \frac{(N+2)}{4(N+8)^2} \epsilon^2 \ . 
\end{align}

\begin{figure}[htbp]
	\begin{center}
		\includegraphics[width=12.0cm,clip]{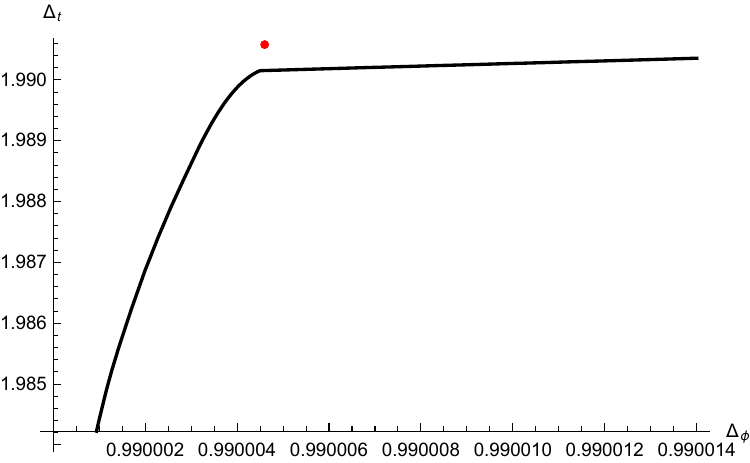}
	\end{center}
	\caption{Unitarity bound for $\Delta_{t}$ in $O(d)$ symmetric conformal field theory in $d=3.98$ dimensions. The red dot represents the dipolar fixed point (in $\epsilon$ expansion).}
	\label{fig:1}
\end{figure}

We can also systematically investigate the scaling dimensions as well as the unitarity bound of the critical $O(N)$ models by using the numerical conformal bootstrap. We show the bound of the scaling dimensions of $\Delta_t$ as a function of $\Delta_{\phi}$ in $O(d)$ model in $d=3.98$ dimensions in Figure 1 by dimensionally continuing the parameter $d$ and $N$.\footnote{We used cboot \cite{cboot} with SDPB \cite{Simmons-Duffin:2015qma} to generate the plot.} It is interesting to observe that within $\epsilon$ expansions, the scaling dimensions of the dipolar fixed point computed by Aharony and Fisher violates the bootstrap bound.
 Of course, this is not a contradiction because the dipolar fixed point does not possess conformal invariance nor reflection positivity, but it is indicative that in a real experiment, we might obtain the number that violates the conformal bootstrap bound, which could result from scale but non-conformal interactions.

After investigating the functional renormalization group approach to the dipolar fixed point, in section 2.3 we will come back to the comparison with bootstrap bond for the Heisenberg model in three dimensions.

\subsection{Wetterich version}
In the following, we would like to study the functional renormalization group approaches to study the dipolar fixed point. We begin our studies with the local potential approximation of the Wetterich equation. The schematic form of the Wetterich equation is
\begin{align}
k \partial_k \Gamma = \frac{1}{2} \mathrm{Tr}\left( \partial_k R_k (\partial_\phi^2 \Gamma + R_k)^{-1} \right) \ ,
\end{align}
where $R_k$ is the regularization functional and we will often use the Litim (or optimal) regulator $R_k = (k^2-p^2)\theta(k^2 -p^2)$ \cite{Litim:2001fd}. 

 Within the local potential approximation, the effective action for the dipolar magnet is truncated as
\begin{align}
\Gamma = \int d^d x \left( \frac{1}{2}\partial_\mu \phi_\nu \partial_\mu \phi_\nu + \xi (\partial_\mu \phi_\mu)^2 + V(\phi_\mu^2) \right) \ .
\end{align}
We assume that $\xi = \infty$ is a fixed point under the renormalization group flow and we do not consider its renormalization as can be justified in the Lagrange multiplier formulation. 

Noting that the inverse of the kinetic term $(p^2 \delta_{\mu\nu} + 2\xi p_\mu p_\nu)^{-1}$ at $\xi = \infty$ is formally given by the Landau gauge propagator $ \frac{\delta_{\mu\nu} -\frac{p_\mu p_\nu}{p^2}}{p^2} = \frac{1}{p^2} P_{\mu\nu} $ with the projector $P_{\mu\nu}$, the Wetterich equation with the local potential approximation becomes
\begin{align}
k\partial_k V = \int \frac{d^dp}{(2\pi)^d} \partial_k R_k P_{\mu\nu} (p^2 \delta_{\nu\mu} + 2(V'\delta_{\nu \rho} + 2V''\phi_\nu \phi_\rho) P_{\rho\mu}) + R_k P_{\nu\mu} )^{-1} \ .  \label{FRGeq}
\end{align}
With the Litim type regulator, the integration over $p$ can be formerly performed
\begin{align}
k\partial_k V =  k^{d+1} \mu_d \langle P_{\mu\nu} (k^2 \delta_{\nu\mu} + 2(V' \delta_{\nu \rho} + 2V''\phi_\nu \phi_\rho) P_{\rho\mu}))^{-1} \rangle_n \ . 
\end{align}
where we still have to evaluate the angular average of the projectors $P_\mu = \delta_{\mu\nu} - \frac{p_{\mu}p_\nu}{p^2}$. For example
\begin{align}
\langle \frac{p_\mu p_\nu}{p^2} \rangle_n = \frac{1}{d} \delta_{\mu\nu} \cr
\langle \frac{p_\mu p_\nu p_\rho p_\sigma}{p^4} \rangle_n = \frac{1}{d(d+2)}(\delta_{\mu\nu} \delta_{\rho\sigma}+ \delta_{\mu\rho}\delta_{\nu\sigma} + \delta_{\mu \sigma} \delta_{\nu \rho}) \cr
\langle \frac{p_\mu p_\nu p_\rho p_\sigma p_{\alpha} p_\beta}{p^6} \rangle  = \frac{1}{d(d+2)(d+4)} (\delta_{\mu\nu}\delta_{\rho\sigma} \delta_{\alpha\beta} + 14 \text{terms}) \ . 
\end{align}
Since it is in the denominator with a non-commuting matrix $\phi_{\mu}\phi_\nu$, the explicit evaluation further is non-trivial. We can, however, always expand the denominator in perturbation theory as we will see.

As our first study, we show how to reproduce the earlier results in $\epsilon$ expansions in $d=4-\epsilon$ dimensions.
For this purpose, we truncate the effective action 
\begin{align}
 V= t \phi_\mu^2 + \lambda (\phi_\mu^2)^2 \ 
\end{align}
and work in perturbation theory with respect to $\lambda$ (and $t$). 

Within the perturbation theory, one can expand the matrix in the denominator and evaluate the angular average up to $\phi^4$. 
The beta function is obtained as
\begin{align}
\dot{t} &= -2t - (2(d-1) + 4 - \frac{4}{d})\mu_d 2 \lambda + 2(2(d-1)+4-\frac{4}{d})\mu_d 4\lambda t + \cdots \cr
\dot{\lambda} &= -\epsilon \lambda + 4 \cdot 4 \lambda^2 \mu_d \left(d+7 -\frac{12}{d} + \frac{12}{d(d+2)} \right) + \cdots \ 
\end{align}
with the fixed point $\lambda_* = \frac{\epsilon}{4\cdot 34 \mu_d} + O(\epsilon)^2$. 
The critical exponent $y_t = d- \Delta_t$ can be computed as
\begin{align}
y_t = 2-\frac{9}{17}\epsilon +O(\epsilon^2) \ 
\end{align}
by linearizing the beta functions at the fixed point and diagonalizing the Hessian matrix $\partial_a \beta^b$.
This reproduces the result by  Aharony and Fisher \cite{AF}. 

In principle, we may study non-perturbative fixed points in $d=3$ dimensions within local potential approximation. Here, we just present one example of (uncontrolled) truncation at the next order in the space of coupling constants. We truncate the effective action 
\begin{align}
 V= t \phi_\mu^2 + \lambda (\phi_\mu^2)^2 + g(\phi_\mu^2)^3 \ 
\end{align}
and demand vanishing of beta-functions of $t$, $\lambda$ and $g$. We also neglect the anomalous dimensions of $\phi$.\footnote{In $\epsilon$ expansion, it is fixed by the momentum-dependent wavefunction renormalization of $O(\epsilon^2)$.} Explicitly we have
\begin{align}
\dot{t} &= -2t - \frac{(2(d-1) + 4 - \frac{4}{d})\mu_d 2 \lambda}{(1+2t)^2} \cr
\dot{\lambda} &= -(4-d) \lambda + \frac{4 \cdot 4 \lambda^2 \mu_d \left(d+7 -\frac{12}{d} + \frac{12}{d(d+2)} \right)}{(1+2t)^3} -  \mu_d \frac{(d-1)+(4-\frac{4}{d})}{(1+2t)^2 }6g  \cr
\dot{g} & = -(6-2d) g + 48 \mu_d g \lambda  \frac{d-1 + 6(1-\frac{1}{d})+8(1-\frac{2}{d}+\frac{3}{d(d+2)})}{(1+2t)^3} \cr
 &-64\mu_d\lambda^3\frac{d-1+6(1-\frac{1}{d})+12(1-\frac{2}{d}+\frac{3}{d(d+2)}) + 8(1-\frac{3}{d}+\frac{9}{d(d+2)} - \frac{15}{d(d+2)(d+4)}) )}{(1+2t)^4}  \ . \label{full}
\end{align}
(Here we have omitted some terms that are higher orders in $\epsilon$ expansions.)
Substituting $d=3$ and linearizing the renormalization group equation around the fixed point, we obtain the lowest renormalization group eigenvalue as
\begin{align}
y_t = 1.529 \ .
\end{align}

In comparison, let us quote the lowest renormalization group eigenvalue in $O(3)$ model in $d=3$ dimensions with the same local potential approximation. It is given by 
\begin{align}
y_t = 1.553 \ .
\end{align}
Note that the scaling dimension $\Delta_t$ obtained here is larger in the dipolar fixed point than the Heisenberg fixed point, which seems consistent with the perturbation theory.\footnote{We cannot trust the actual number very much. For example, the conformal bootstrap suggests that $y_t=1.406$ for the $O(3)$ model in $d=3$ dimensions.}

We could actually write down the full functional form of the renormalization group equation in $d=3$ dimensions.\footnote{The following observation was first suggested by K.~Fukushima.} We first evaluate the effective propagator in the Wetterich equation:
\begin{align}
G_{\mu\nu} = \tilde{A} P_{\mu\nu} + \tilde{C} P_{\mu\alpha} \phi_\alpha P_{\nu\beta} \phi_\beta 
\end{align}
where
\begin{align}
\tilde{A} &= \frac{1}{p^2 + 2V'} = \frac{1}{\bar{p}^2} \cr
\tilde{C} &  = -\frac{p^2}{\bar{p}^2} \frac{4V''}{p^2(\bar{p}^2 + 4V''\phi_\mu^2) - 4V''(p_\mu\phi_\mu)^2} \ .
\end{align}
Let us now perform the angular average of $p$ integration in the right-hand side of the Wetterich equation in $d=3$. It is effectively given by
\begin{align}
&\frac{2}{\bar{p}^2} + \frac{1}{2} \int_{-1}^1 d (\cos\theta) \frac{-p^2}{\bar{p}^2} \frac{4V'' \phi_\mu^2 (1-\cos\theta^2)}{p^2 (\bar{p}^2 + 4 V'' \phi_\mu^2) - 4 V''  p^2 \phi_\mu^2 \cos^2\theta} \ \cr
=&  \frac{2}{\bar{p}^2} - \frac{p^2}{\bar{p}^2}\frac{2V''\phi_\mu^2}{p^2 \bar{p}^2 + 4 V'' p^2 \phi_\mu^2} \int_{-1}^1 dx \frac{1-x^2}{1-\frac{4V''p^2\phi_\mu^2}{p^2\bar{p^2}+ 4V'' p^2 \phi_\mu^2} x^2} \cr
=&  \frac{2}{\bar{p}^2} - \frac{p^2}{\bar{p}^2}\frac{2V''\phi_\mu^2}{p^2 \bar{p}^2 + 4 V'' p^2 \phi_\mu^2} \frac{2 a + \frac{-1+a^2}{2} \log \left(\frac{1+a}{1-a}\right)^2}{a^3} \ , 
\end{align}
where $a^2 = \frac{4V''p^2\phi_\mu^2}{p^2\bar{p^2}+ 4V'' p^2 \phi_\mu^2}$. By performing the polar integration with the optimal regulator, we get
\begin{align}
k\partial_k V &=   \frac{2}{\bar{k}^2} - \frac{k^2}{\bar{k}^2}\frac{2V''\phi_\mu^2}{k^2 \bar{k}^2 + 4 V'' k^2 \phi_\mu^2} \frac{2 \bar{a} + \frac{-1+\bar{a}^2}{2} \log \left(\frac{1+\bar{a}}{1-\bar{a}}\right)^2}{\bar{a}^3} \cr
&=\frac{2}{\bar{k}^2} - \frac{k^2}{\bar{k}^2}\frac{2V''\phi_\mu^2}{k^2 \bar{k}^2 + 4 V'' k^2 \phi_\mu^2} \sum_{n=1}\frac{4\bar{a}^{2n-2}}{4n^2-1}  \  
\end{align}
with $\bar{k}^2 = k^2 +2 V'$ and  $\bar{a}^2 = \frac{4V''k^2\phi_\mu^2}{k^2\bar{k^2}+ 4V'' k^2 \phi_\mu^2}$.
One can check that it reproduces the beta functions we obtained perturbatively above. 

\subsection{LPA' and more results}
One may incorporate the effect of the anomalous dimensions within the functional renormalization group approach. We do not attempt the evaluation of the wavefunction renormalization in a self-consistent manner, which is technically more involved. Here we take the approach called LPA' and put the effect of the wave function renormalization ``by hand". In this approach, the net effect of the wavefunction renormalization is given by replacing \eqref{FRGeq} with
\begin{align}
k\partial_k V = \partial_k (k^{d+2} Z_k) \mu_d \langle P_{\mu\nu} (Z_k k^2 \delta_{\nu\mu} + (2V' \delta_{\nu\rho} + 2 V'' \phi_\nu \phi_\rho) P_{\rho\mu}))^{-1}\rangle_n \ ,
\end{align}
where we {\it assume} $Z_k \sim k^{-2 \gamma_{\phi}}$ with $\gamma_\phi$ being the anomalous dimension of $\phi_i$ that can be computed separately.  Within the LPA' approach, where we put the value of $\gamma_\phi$ by hand, the resulting renormalization group equations are almost the same as \eqref{full} except that the coefficient of the first term is modified: for $g_n \phi^{n}$ coupling, we replace $-(n - \frac{dn}{2} + d)g_n$ with $ -(n - \frac{dn}{2} + d- n \gamma_\phi)g_n$. 

The values of $\gamma_\phi$ can be taken from the perturbative computations based on the epsilon expansions (or any other methods). At $d=3$, we have $\gamma_\phi \sim 0.01(1)$, which only gives a tiny modification of the (lowest) renormalization group eigenvalues $y_t$ (of order $\gamma_\phi$: see Figure 2 below).

We report the evaluation of $\Delta_t = 3-y_t$ as a function of $\Delta_\phi = \frac{1}{2} + \gamma_\phi$ in the Aharony-Fisher model (in $d=3$) within the LPA' approximation by changing the truncation of the potential in Figure 2.
To quote some numbers here, if we truncate the potential up to $\phi^6$, we 
 obtain $y_t = 1.508$ or if we truncate the potential up to $\phi^{16}$ we obtain $y_t =  1.33$ (at $\gamma_\phi= 0.02$). The small dependence on $\gamma_\phi$ can be extrapolated from Figure 2.

 The prediction from $y_t$ by increasing the truncation order 
 of the potential seems to converge rapidly, but this does not mean that we can trust the actual number we have obtained. While we cannot estimate the systematic error in the Ahanorny-Fisher fixed point, with the same truncation, we obtain $y_t = 1.31$ in the $O(3)$ model, whose accurate value should be $y_t = 1.406$. It is therefore expected that the systematic error of our prediction of $y_t$ could be as large as $0.1$ irrespective of the convergence of the polynomial truncations within the LPA'. See also \cite{Dupuis:2020fhh}\cite{Murgana:2023xrq} for similar comparisons in $O(N)$ models. 
 Note that the effect of the truncation (of the other terms we neglect in LPA') seems much more severe than the effect of the anomalous dimensions $\gamma_{\phi}$. 


\begin{figure}[htbp]
	\begin{center}
		\includegraphics[width=12.0cm,clip]
  {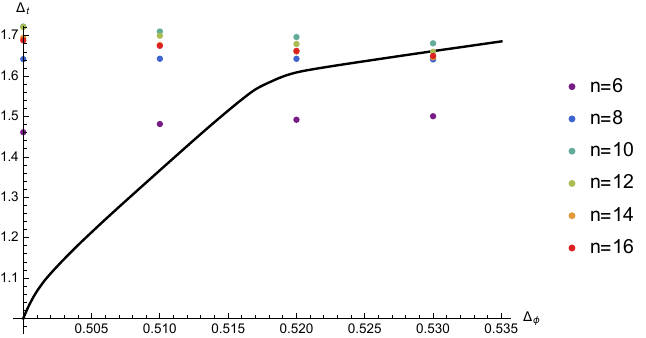}
	\end{center}
	\caption{The critical exponents obtained from LPA' truncation of the functional renormalization group in three dimensions. For comparison, we showed the conformal bootstrap bound on the Heisenberg model as a black curve.}
	\label{fig:2}
\end{figure}

Let us finally quote the predictions of $y_t$ (or $\Delta_t = 3-y_t)$ from various other approaches. The three-loop computations of the renormalization group directly in three dimensions  \cite{Kudlis} gave $\Delta_\phi = 0.5165(40)$, $\Delta_0 = 1.576(10)$. The experimental values (more than forty years ago) in EuO and EuS gave $\Delta_0 = 1.58(5)$ and $1.59(5)$ respectively \cite{exp}. 

\subsection{Polchinski version}
Next, let us study the local potential approximation of the Polchinski equation as another functional renormalization group approach to the dipolar fixed point. 
The schematic form of the Polchinski equation for the Aharony-Fisher model is given by
\begin{align}
\dot{S} =  -\frac{\delta S}{\delta \phi_\mu(p)} P_{\mu\nu} \frac{\delta S}{\delta \phi_\nu(-p)} + \mathrm{Tr} P_{\mu \nu}\frac{\delta^2 S}{\delta \phi_\mu(p) \phi_\nu(-p)} \ .  \label{Polchi}
\end{align}
One apparent advantage of the Polchinski equation (compared with the Wetterich equation) is the absence of the denominator.

The important difference compared with the standard scalar $\phi^4$ theory is to keep the projector $P_{\mu\nu} = \delta_{\mu\nu} - \frac{p_\mu p_\nu}{p^2}$ in the interaction vertex even in the local potential approximation. We also perform the angular average when we take the trace in the second term of \eqref{Polchi}, but we do not perform the average in the first term.
This makes the solution of the Polchinski equation much more complicated, but it is necessary even in the perturbation theory. 

As our first application, let us study a perturbative fixed point in $d=4-\epsilon$ dimensions. 
In order to make the renormalization group equation closed within the perturbation theory, we make the ansatz:\footnote{We need the six-point vertex to reproduce the standard $\epsilon$ expansions in standard Wilson-Fisher fixed point from the Polchinski equation.}
\begin{align}
V(\phi) =  t \phi_\mu P_{\mu\nu} \phi_\nu + \lambda \phi_\mu \phi_\mu \phi_\nu \phi_\nu + g \phi_\mu \phi_\mu \phi_\nu P_{\nu\sigma} \phi_\sigma \phi_\rho \phi_\rho \ . 
\end{align}
Note that the six-point vertex has a specific projector.\footnote{Note that if the projector is connected to only one $\phi$ (i.e. in $t$ term), it does nothing because the external line is always transverse. On the other hand, if the projector connects more fields (i.e. in $g$ term) then it makes a difference.} The fixed point equation for $g$ at the lowest order becomes
\begin{align}
0 = -16\lambda^2 \phi_\mu \phi_\mu \phi_\nu P_{\nu\sigma} \phi_\sigma \phi_\rho \phi_\rho  - 2 g \phi_\mu \phi_\mu \phi_\nu P_{\nu\sigma} \phi_\sigma \phi_\rho \phi_\rho \  ,
\end{align}
which indeed shows the necessity of the projector.

Similarly for $t$, we have
\begin{align}
0 = 2t \phi_\mu \phi_\mu + (2(d-1) + 4(1-\frac{1}{d})) \lambda \phi_\mu \phi_\mu  - 4 t^2 \phi_\mu P_{\mu\nu} \phi_\nu \ ,
\end{align}
We should note that for the two-point vertex, there is no distinction between $\phi_\mu \phi_\mu $ and $\phi_\mu P_{\mu\nu} \phi_\nu$, so we can combine all these terms and demand vanishing of the coefficient. 

The fixed point equation for $\lambda$ has two contributions. One is the one-particle reducible one
\begin{align}
 -16 t \lambda (\phi_\mu \phi_\mu \phi_\nu) P_{\nu \sigma} \phi_{\sigma}  +(2g(d-1) + 4g(1-\frac{1}{d}))(\phi_\mu \phi_\mu \phi_\nu) P_{\nu \sigma} \phi_{\sigma} 
\end{align}
and the other is the one-particle irreducible one
\begin{align}
 g(d+7-\frac{12}{d}+\frac{12}{d(d+1)} ) \phi_\mu \phi_\mu \phi_\rho \phi_\rho \  . \end{align}
At the fixed point, we see that the one-particle reducible contribution cancels with each other and we have the fixed point equation for $\lambda$: 
\begin{align}
\dot{\lambda} = \epsilon \lambda - 8 \lambda^2 (d+7-\frac{12}{d}+\frac{12}{d(d+2)})
\end{align}
with the fixed point value of $\lambda_* =\frac{\epsilon}{4\cdot 17} $ (and $g_* = -8 \lambda_*^2$ and $t_* = -\frac{9}{2} \lambda_*$). We can compute the RG eigenvalues, and we obtain $y_t = 2-\frac{9}{17} \epsilon $ correctly.

Our original hope was that the Polchinski equation may work better to study the non-perturbative renormalization group fixed point in the Aharony-Fisher model (at least within the local potential approximation) because of the absence of the denominator. Unfortunately, it may not be that simple. Due to the existence of the projector, we may have to introduce more and more terms 
\begin{align}
V = t \phi^2 + \lambda_0 \phi^2 \phi^2 + \lambda_1 \phi \phi P \phi \phi + g_0 \phi^2 \phi^2 \phi^2 + g_1 \phi^2 \phi P \phi \phi^2 + g_2 \phi \phi P \phi \phi P \phi \phi + \cdots \  
\end{align}
to write down the effective action. It is not obvious how to truncate such potentials or make any non-perturbative ansatz that is closed under the renormalization group flow.

\section{$d=2$ and multicritical points}
The physical motivation of the dipolar fixed point mainly resides in $d=3$ dimensions, but we may be able to find a non-trivial fixed point also in $d=2$ dimensions. Note that the ordinary $O(2)$ model does not show spontaneous symmetry breaking in $d=2$ dimensions due to the Coleman-Mermin-Wagner theorem, but it does not apply to the Aharony-Fisher model because the global symmetry is mixed with the rotational symmetry.

In two dimensions, the transverse vector can be replaced by a scalar with a (gauged) shift symmetry:
\begin{align}
\phi_\mu = \epsilon_{\mu\nu} \partial_\nu \varphi \ .
\end{align}
with $\varphi$, the Landau-Ginzburg effective action for the Aharony-Fisher model can be represented as
\begin{align}
S = \int d^2x\left( \partial^2 \varphi \partial^2 \varphi + V(\partial_\mu \varphi \partial_\mu\varphi)  + \cdots \right) \ . 
\end{align}
When $V = 0$, the theory is globally conformal invariant but not Virasoro invariant \cite{Nakayama:2016dby}\cite{Nakayama:2019xzz}. It is not obvious if non-trivial multi-critical fixed points with $V\neq 0$ admit (global) conformal invariance. Presumably, they do not,\footnote{In \cite{Gimenez-Grau:2023lpz}, it is conjectured that an interacting fixed point with shift symmetry (like the one here) is only scale invariant without conformal invariance based on the genericity argument.} but in either case, we may find these non-trivial renormalization group fixed points.

While we may study non-trivial fixed points from the functional renormalization group directly in the original variable $\phi_\mu$ which is transverse,  we may also study them from the new variable $\varphi$ without any constraint.
In the local potential approximation with the optimal regulator, the Wetterich equation of this model is given by
\begin{align}
k \partial_k V = k^{d+1} \langle \frac{1}{k^2 + 2V'(\partial_\mu \phi \partial_\mu \phi ) + 4k^{-2} V''(\partial_\mu \phi \partial_\mu \phi ) \partial_\rho\phi \partial_\sigma \phi k_\rho k_\sigma } \rangle_n \ . 
\end{align}
This is similar, but slightly different from the equations discussed before in terms of $\phi_\mu$. 
Since the truncation we are using here is equally uncontrolled, we cannot say which would give a more reasonable result. Note that here again we have to expand the denominator to evaluate the angular average, and the computational difficulty has not been alleviated. 

Actually, we can perform the angular average in $d=2$. It is given by
\begin{align}
k\partial_k V &= k^{3} \frac{1}{2\pi} \int_{-\pi}^{\pi} d\theta \frac{1}{k^2 + 2V' + 4(\partial_\mu \varphi)^2 V''  \cos^2\theta}  \cr
&= k^3 \frac{1}{\sqrt{k^2 + 2V' }\sqrt{k^2 + 2V' + 4 (\partial_\mu \varphi)^2 V'' }}
\end{align}
It may give a starting point to study the functional analysis of the fixed point potential $V$.

As in conformal minimal models in $d=2$ dimensions, we expect that the model admits (infinitely many) multi-critical fixed points by fine-tuning $V$. They can be regarded as scale but non-conformal analogue of minimal models. It would be very interesting to study their properties and the renormalization group flow among them.

\section{Discussions}
In this paper, we have presented our first attempt to use the functional renormalization group method to study the critical exponents of the dipolar fixed point. There are a couple of directions to be explored. One is to do a systematic search for the non-perturbative fixed point without doing a brute-force truncation of the potential even within the local potential approximation. 

Another important direction is to introduce the effect of the wavefunction renormalization to compute the critical exponent $\eta$. Even in perturbation theory, it is non-trivial to compute $\eta$ in the functional renormalization group approach \cite{Papenbrock:1994kf}\cite{ODwyer:2007brp}\cite{Codello:2013bra}, and it requires to compute the field-dependent wavefunction renormalization at the dipolar fixed point. In the perturbative functional renormalization group, $\eta$ can be related to the 
 terms  such as $Z_{\phi^2}(\phi^2) \partial_\mu \phi_\nu \partial_\mu \phi_\nu$ in the one-loop effective action. Now $Z_{\phi^2}$ itself is of order $\lambda^2$ in the one-loop integral of the bare Lagrangian, so $\eta$ is of order $\lambda^2$ corresponding to the effective two-loop integral.
It is crucial to obtain $\eta$ non-perturbatively in $d=3$ dimensions in order to see if the dipolar fixed point really violates the conformal bootstrap bound for the $O(3)$ models in $d=3$ dimensions. 

\section*{Acknowledgements}
The author thanks S.~Yabunaka for the discussions on (perturbative) functional renormalization group. He acknowledges the Yukawa Institute for Theoretical Physics at Kyoto University. This work is based on the author's talk at YITP-W-20-09 ``10th International Conference on
Exact Renormalization Group 2020"  and discussions there were useful to complete this work. In particular, he is grateful to K.~Fukushima for his valuable comments and encouragement. He would like to thank A.~Gimenez-Grau and S.~Rychkov for the subsequent collaboration. This work was in part supported by JSPS KAKENHI Grant Number 17K14301.

\end{document}